\documentstyle[PASJadd, epsf]{PASJ95}
\draft

\begin{document}
\setcounter{page}{1}

\title{An Extremely Red Nucleus
in an Absorbed QSO at $z=0.65$}

\author{
Masayuki {\sc Akiyama}\thanks{Visiting astronomer of the University of
Hawaii $88^{\prime\prime}$ telescope.} \\
{\it Subaru Telescope, National Astronomical Observatory of Japan,}\\
{\it 650 North A'ohoku Place, Hilo, HI, 96720, U.S.A}\\
{\it E-mail(MA): akiyama@naoj.org} \\
and\\
Kouji {\sc Ohta}\footnotemark[1] \\
{\it Department of Astronomy, Faculty of Science,
Kyoto University, Kyoto 606-8502} \\
{\it E-mail(KO): ohta@kusastro.kyoto-u.ac.jp} \\
}

\abst{
The results of $K$-band-imaging observations of a candidate
of an absorbed QSO at $z=0.653$, AX~J131831+3341, are presented.
The $B-K$ color of the object is 4.85 mag, which is
much redder than optically-selected QSOs. 
The $K$-band image shows nuclear and extended components,
the same as in the optical $V$-, $R$-, and $I$-band images.
The nuclear component ($I-K = 4.29$ mag) is much redder than
the power-law models with energy indices of 0 to $-1.0$,
which well reproduce the
$V-R$ and $R-I$ optical colors of the nuclear component.
A heavily absorbed ($A_V \sim 3$ mag) nucleus may emerge in the $K$-band,
while optical light may originate from scattered nuclear light.
The $I-K$ color of the extended component is 2.2 mag, which is
consistent with the post-starburst nature
of the host galaxy, which is also suggested from the $V-R$ and $R-I$ colors
of the extended component.
}

\kword{galaxies: active --- galaxies: individual (AX~J131831+3341)
--- galaxies: photometry --- quasars}

\maketitle
\thispagestyle{headings}

\section{Introduction}

X-ray-selected absorbed radio-quiet QSOs are a new population of AGNs
which have appeared in recent deep X-ray surveys (Almaini et al. 1995;
Ohta et al. 1996; Kim, Elvis 1999; Fiore et al. 1999).
Although their X-ray luminosities reach those of QSOs,
the absence of a strong broad Balmer emission line
or their large Balmer decrements of broad
emission lines
suggests the existence of
strong absorption to their nuclei.
These objects are considered to be the long-sought
luminous cousins of the type-2 Seyfert galaxies, in other words, 
an absorbed
version of broad-line non-absorbed QSOs.
Their optical and near-infrared colors
are normally redder than optically-selected non-absorbed QSOs
(Georgantopoulos et al. 1999; Nakanishi et al. 2000).
The origin of their red colors is thought to be
an absorbed nuclear continuum or that from a host galaxy of the QSO.

AX~J131831+3341 is another candidate of an absorbed radio-quiet
QSO at a redshift of 0.653 found during the course of
the optical identification of hard X-ray sources
in the ASCA Large Sky Survey
(LSS; Akiyama et al.\ 2000a, hereafter Paper I).
Its X-ray luminosity is estimated to be $\sim 10^{45}$ erg
s$^{-1}$.
Deep $R$- and $V$- band images reveal the presence of a
point-like nucleus and an extended component (Akiyama et al.\ 2000b,
hereafter Paper II).
The $R$-band magnitude of the nuclear component
is much fainter than that expected from the
observed X-ray flux with optical to X-ray flux ratios of
X-ray-selected broad-line QSOs.
The optical faintness suggests that the nucleus
is absorbed with $A_V$ of larger than 3 mag.
However, the optical $V-R$ and $R-I$ colors of the nuclear component are
as blue as a power-law model with an energy index of $\alpha=-1$
($f_{\nu} \propto \nu^{\alpha}$).
Although the colors are slightly redder than the averaged color
of QSOs ($\alpha=-0.3$--$-0.5$, Francis et al. 1996),
an absorption of only $A_V \leq 1$ mag is required to reproduce
the colors.
Thus, the optical colors conflict with the amount of
absorption to the nucleus, estimated from the optical faintness.
Another important property of this object is that a broad
Mg {\scriptsize II} 2800 {\AA} emission line is seen in the
optical spectrum (Paper I).
Therefore, in order to explain all of these properties
simultaneously, the origin of the optical nuclear component is
strongly suggested to be a
scattered nuclear light and the nuclear emission is heavily
absorbed ($A_V \geq 3$ mag), though there is a possibility that the
nuclear component is a slightly absorbed ($A_V \sim 1$ mag) nucleus, if
its intrinsic X-ray to optical flux ratio is the largest among
X-ray-selected broad-line QSOs (Paper II).
The scattered nuclear light should contain a broad H$\beta$ emission
line as well as the broad Mg {\scriptsize II} 2800 {\AA}
emission line.  The absence of the broad H$\beta$ emission line
in the optical spectrum (Paper I) can be caused by 
contamination of the host galaxy continuum, because the extended
component contributes to 50\% of the $I$-band light in
a $1.\!^{\prime\prime}2$ circular aperture, which corresponds
to a slit width of $1.\!^{\prime\prime}2$.
In order to investigate the near-infrared to optical spectral
energy distribution of the object, and to unveil the
origin of the near-infrared to optical continuum component,
we conducted a $K$-band imaging observation.
Throughout this paper we use $q_0=0.5$ and $H_0=50$ km s$^{-1}$
Mpc$^{-1}$.

\section{Observations}

The $K$-band imaging observations were made with a QUick InfraRed Camera (QUIRC)
with a 1024 $\times$ 1024 HgCdTe Astronomical Wide Area Infrared Imaging (HAWAII)
array attached to the $f/$10 focus of the University of Hawaii $88^{\prime\prime}$
telescope on 2000 March 21 and 22.
Sixteen and twenty-eight frames with an exposure time of 180 s
were taken on March 21 and 22, respectively, with small offsets of the
telescope pointing.
The pixel scale of the camera was $0.\!^{\prime\prime}189$ pixel$^{-1}$.
The seeing condition during the observation was
FWHM of stars of $\sim 0.\!^{\prime\prime}9$ and
$\sim 1.\!^{\prime\prime}2$ in the first and second nights, respectively.
During the observing run, UKIRT faint standard stars
(FS 121, FS 124, FS 129, FS 126, FS 132, FS 137, and FS 135)
were observed for a photometric calibration.
For each star, we took at least 3 images, while changing the position
of the star on the detector to reduce any systematic errors.

The data reduction was performed as follows:
at first, we made sky-background frames by 
stacking object frames of each night
without a shift.
The sky-background frame of the same night
was subtracted from each object frame.
Next, flat-fielding was performed
with a dome-flat frame.
Finally, after correcting the offset of each object frame,
we combined 11 and 19 object frames 
taken under a good seeing condition
in the first and second night, respectively.
Sky-subtraction and flat-fielding were also performed for standard star
frames in the same manner, 
though we did not shift and combine the standard star
frames.
The counts of the standard star in each frame were measured individually
by using a growth curve-fitting method for each star.
Based on the scatters of the count rate to the magnitude conversion factors
derived from the observed UKIRT faint standard stars,
we estimated the
uncertainties of the photometric calibrations as
0.05 mag for the first night data and 0.1 mag for the second night data.
The uncertainty includes those of the flat-fielding.

\section{Results}
\subsection{Morphology and Total Magnitude}

The $K$-band image made from the first and the second night
data is shown in figure 1.
AX~J131831+3341 shows a point-like nucleus and an extended component,
just as in the optical image in Paper II.
The profile of the nuclear component is consistent with
that of stars in the image.
The standard deviation in the sky region of the image
is 7 counts per pixel, which corresponds to 22.2
mag arcsec$^{-2}$ in the $K$ band.
In figure 1, we show a sky-subtracted image of the object
with the surface brightness ranging
from $-3$ times the standard deviation to $+10$ times
the standard deviation as a gray scale with a linear scale.


We measured the total magnitude of AX~J131831+3341
in a $26.\!^{\prime\prime}4$ aperture centered on the nucleus,
excluding objects around AX~J131831+3341 and a
knot in the northwestern direction, as we did for the optical images
in Paper II.
The resulting $K$-band magnitudes are 16.38$\pm$0.14 mag
and 16.28$\pm$0.16 mag from the first and the
second night data, respectively.
The uncertainty of the magnitude
includes that of the photometric calibration and
that of the sky determination (0.13 mag).
The two magnitudes agree with each other within the uncertainty.
The resulting $B-K$ color of AX~J131831+3341 is 4.95$\pm$0.17 mag;
it is as red as other X-ray-selected absorbed QSOs 
($B-K=5.4$ mag for RX~J13334+0001 at $z=2.35$,
Georgantopoulos et al. 1999; $B-K=5.3$ mag for AX~J08494+4454 at
$z=0.9$, Akiyama et al. in preparation).
In the next section, we estimate the magnitudes and colors of
the nuclear and the extended components separately.


\subsection{Colors and Magnitudes of the Extended and the Nuclear Components}

In order to evaluate the colors of the extended component,
we measured the magnitude in regions A and B
($4.\!^{\prime\prime}8 \times 3.\!^{\prime\prime}6$ and
$2.\!^{\prime\prime}4 \times 3.\!^{\prime\prime}6$) shown in figure 1.
These regions are the same as those used for the optical images in Paper II.
Because the data taken during the first night have a smaller seeing
size, and the size is not so much different from that of the optical images
($0.\!^{\prime\prime}7 \sim 0.\!^{\prime\prime}8$),
we use only the first-night data for the color measurements.
The magnitude in region A (B) was measured to be 18.56 (18.43) mag.
The uncertainty of the photometry is estimated to be
0.11 mag, which includes the uncertainties of the photometric zero point
and the sky determination (0.1 mag).
Thus, the color in region A (B) is $I-K$ of 2.54 (2.49) $\pm$ 0.19 mag.


To derive the $K$-band magnitude of the nuclear component,
we deconvolved the nuclear and extended components,
following the same method as used in the deconvolution of
the $V$-, $R$-, and $I$-band images (Paper II).
Since the object has an asymmetric, complex feature,
in order to obtain a rather symmetric and smooth brightness
distribution, we, at first,
modeled the surface-brightness distribution of the
galaxy by fitting ellipses, of which the centers, position angles,
ellipticities, and surface brightnesses were free parameters, to the
isophotes of the object.
We adopted the same sampling steps as those used for the optical images;
the step size was taken to be one twelfth of the semi-major length
of each ellipse.
The constructed model and the residual image after
subtracting the model from the original image are shown in figures 1b and 1c.
(All images in figure 1 are shown in the same count-rate range.)
The model well describes the overall shape of the original image.
The surface-brightness profiles were derived as sections of the
model image along the major-axis (position angle of $114^{\circ}$),
and are shown in figure 2 by the thick solid lines.
We fit the profiles with a model consisting of a nuclear component
(a point source) and an exponential disk component.
The profile of the point-spread function
was determined by applying the same profile measuring method
as that applied for the object
to the three stars, one of which was the reference star for the frame alignment.
We assumed that the scale length of the exponential disk
in the $K$ band was the same as that in the $R$-band image, and
left the normalizations of the point source and the exponential disk
as free parameters.
The best-fit normalizations were
determined from profiles between $2^{\prime\prime}$ and $3^{\prime\prime}$
where the exponential disk component dominates the profile
and within $1^{\prime\prime}$ where the nuclear component dominates
the profile.
The summed profiles of the best-fit models are shown by the thick dashed
lines in figure 2.
The model well describes the profile of the object in both directions.
The resulting magnitude of the nuclear component is 17.48 mag and
the central surface brightness of the extended component is
20.00 (19.82) mag arcsec$^{-2}$ in the southeastern (northwestern)
profile.
We plot models whose normalizations of the
nuclear component and the extended component are changed by
$\pm0.2$ mag by the thin solid lines.
Because most of the observed profiles run between these thin profiles,
the uncertainties of the parameters are estimated to be less than
$\pm 0.2$ magnitude.
Subtracting the magnitude of the nuclear component from the
total magnitude, we obtain the magnitude of the extended component
as 16.87$\pm$0.25 mag. As a result, the $I-K$ color of the
extended component was calculated to be 2.17$\pm$0.30.

\section{Discussions}
\subsection{Red Nuclear Component}

The $I-K$ and $R-I$ colors of the nuclear component are plotted in
figure 3 (filled square).
The $I-K$ color of the nuclear component is 4.29$\pm$0.25 mag. It is
much redder than those of optically selected QSOs at
redshifts of between 0.2 and 1.5 (open squares in figure 3 from
Elvis et al.\ 1994), and is also much redder than those of
the power-law continuum with energy indices of 0 to $-1.0$ (open pentagons
in figure 3),
which reproduce the $V-R$ and $R-I$ colors of the
nuclear component well with no or a slight extinction.
Since a reddening of the power-law continuum cannot explain the red
color appearing in figure 3, this $I-K$ color can be explained by neither
the scattered nuclear light nor slightly absorbed nuclear emission.
In the $K$ band, a contribution from another continuum emission, which
is at least $\sim$5-times brighter than the optical power-law component
in the rest frame 1.3 $\mu$m, is required.
Candidates of such components are
heavily absorbed direct nuclear emission,
dust thermal emission,
relativistically beamed synchrotron emission,
and host galaxy emission.

One possible origin is that a heavily absorbed red nucleus emerges in 
the $K$-band.
Since the component must dominate the $K$-band flux, but
must not contribute significantly to the $I$-band flux, an absorption larger
 than $A_V = 3$ mag in the rest frame is required.
The amount of the absorption is within the range
derived from the X-ray spectrum of the object ($A_V = 1$ -- 6 mag).
Based on the observed $K$-band magnitude of the nuclear component,
the $A_V$ of 3 mag in the rest frame ($A_K=0.9$ mag in the observed
frame), and the typical energy index of the QSO optical power-law continuum
($\alpha = -0.5$, i.e., $V-K = 2.68$ mag),
the absorption corrected $V$-band magnitude of the nucleus is
expected to be $18.9$ mag.
This leads to an intrinsic absorption-corrected X-ray to optical flux
ratio, $\log (f_{\rm X_{0.3-3.5 keV}}/f_V)$, of $+0.83$.
The value is within the
range of the X-ray to optical flux ratio of
AGN sample of Einstein Medium Sensitivity Survey
[$\log f_{\rm X_{0.3-3.5 keV}}/f_V = -1.0$ -- $+1.4$]
(Stocke et al. 1991).
Therefore, it is plausible that
the nucleus is heavily absorbed ($A_V \sim $ 3 mag) and
emerges only in the $K$ band, while in the optical wavelength
we see the scattered blue nuclear emission.
Comparing the observed $V$-band magnitude of the nuclear component
($V$ = 22.61 mag, Paper II)
with the absorption corrected $V$-band magnitude of the nucleus,
we estimated the fraction of the scattered component as 2\%,
which is similar to that observed in radio galaxies
(Alighieri et al. 1994).

The other possible origin for the red $I-K$ color is
thermal emission by hot dusts in the $K$ band.
The dust thermal emission with a temperature of 1300 -- 1500 K
starts to contribute to the continuum
of QSOs at $1.0 \mu$m (Pier, Krolik 1993; Kobayashi et al.\ 1993).
The $K$-band nuclear component is at least $\sim$5-times brighter than
that expected from the optical power-law continuum component at rest-frame 1.3 $\mu$m.
This value is much larger than the estimated contribution by the dust
thermal emission at 1.3 $\mu$m: less than 1 (Pier, Krolik 1993)
and less than 2 (Kobayashi et al.\ 1993).

For red quasars found in the Parkes flat-spectrum radio
source survey, a relativistically beamed synchrotron component
is proposed as the origin of the red colors
(Serjeant, Rawlings 1995; Srianand, Kembhavi 1997; Francis et al. 2000).
However, because AX~J131831+3341 has a radio-to-X-ray flux ratio similar to
those of radio-quiet AGNs, the existence of such a strong
synchrotron component is very unlikely.

It is possible that a stellar population in the
nuclear region emerges in the $K$ band.
We plot the colors of the spectral
evolution models of galaxies in figure 3. We use two models (Kodama,
Arimoto 1997):
1) An elliptical model (plotted with a solid line) in which star formation
occurs during the first 0.353 Gyr with an initial mass function with a
slope of 1.20 and after that the galaxy evolves passively.
The model parameters well reproduce the reddest and brightest ($M_V
= -23$ mag) class elliptical galaxy in the Coma cluster (Kodama et al.\ 1998).
2) A disk model (plotted with a dotted line) in which star formation
occurs constantly with the same initial mass function as the elliptical
model.
The colors of these models at ages from 0.01 Gyr to 12 Gyr at redshift
of 0.653 are shown as tracks.
We mark the positions at 6 Gyr age with tick marks on the
tracks of both models. The age corresponds to that of the universe
at a redshift of 0.653 under the adopted cosmological parameters
of $q_0 = 0.5$ and $H_0 = 50$ km s$^{-1}$ Mpc$^{-1}$.
If we assume that the stellar population dominates the $K$-band
light, but does not
significantly contribute to the nuclear component in the $I$ band
(and the bluer bands),
the stellar population must have a redder $I-K$ color than the
observed $I-K$ color of the nuclear component.
Without absorption, the reddest model
($I-K$ of 3.4 mag, an elliptical model with an age of 6 -- 12 Gyr)
is still bluer than the $I-K$ color of the nuclear component.
Therefore, to reproduce the red $I-K$ color,
at least absorption with $A_V$ of 1 mag is required.

Such a red optical to near-infrared color and a blue optical color
may be common characteristics of X-ray selected absorbed QSOs at
intermediate redshifts.
Recent observations reveal that
another absorbed QSO at a redshift of 0.9, AX~J08494+4454,
also has a blue optical color
and a red optical to near-infrared color ($R-I$ = 0.67 mag and
$I-K$ = 3.4 mag, Nakanishi et al. 2000; Akiyama et al. in preparation).
Also, a significant fraction of
the optical and near-infrared counterparts of Chandra
hard X-ray sources show similar red near-infrared colors
($I-HK'=4$--$5$ mag) and blue optical colors ($B-I=1$--$2$ mag) (Mushotzky et
al.\ 2000).

\subsection{Nature of the Extended Component}

The extended component is thought to be the host galaxy of
the QSO. The optical absolute magnitude of the component
is similar to those of the brightest host galaxies of QSOs (Paper II).
The $V-R$ and $R-I$ colors of the component are consistent
with a 1 Gyr-old stellar population model without absorption.
If we introduce optical extinction ($A_V=1$--$3$mag) in the host galaxy,
the colors are also explained by a disk model at any ages or
an elliptical at ages of less than 0.1 Gyr (Paper II).

The $I-K$ and $R-I$ colors of the extended component
are plotted in figure 3 (filled circle). The colors
agree with those of the 1 Gyr-old elliptical galaxy model.
If we introduce an optical extinction,
neither a disk model at any ages nor an elliptical model
at ages of less than 0.1 Gyr
can reproduce the $R-I$ and $I-K$ colors.
Therefore the models involving the moderate absorption are rejected.
Since a spectrum of the 1 Gyr-old elliptical model resembles that of a
post-starburst galaxy, these colors
may indicate that the host galaxy is in a post-starburst phase.
The $I-K$ and $R-I$ colors measured in 
region B (filled pentagon) is also consistent with those of 
the 1 Gyr-old elliptical galaxy, 
but the $R-I$ color of region A (filled triangle)
is bluer than that predicted by the model.
Without extinction,
the $R-I$ and $I-K$ colors of region A are not consistent with any
galaxy models at ages less than 6 Gyr, which is a cosmic age
at a redshift of 0.653.
If we introduce an optical extinction in region A,
the colors can be explained by a disk model at any ages
with an absorption of $A_V = 1$ -- $3$ mag
or the elliptical model at ages of less than 0.1 Gyr with
an absorption of $A_V\sim3$ mag.
Therefore, the colors of region A may reflect an 
obscured star-forming region in the host galaxy.

\par
\vspace{1pc}\par
MA and KO appreciate support by K. Nakanishi and staff members of the
 UH observatory during the imaging observations.
This research made use of the NASA/IPAC Extragalactic Database (NED),
which is operated by the Jet Propulsion Laboratory, Caltech,
under a contract with the National Aeronautics and Space Administration.
MA acknowledges support from 
Research Fellowships of the Japan Society for the Promotion of Science
for Young Scientists.
KO's activity is supported by a grand-in-aid from the
Ministry of Education, Science, Sports and Culture (11740123).

\clearpage
\section*{References}
\re
Akiyama, M., Ohta, K., Tamura, N., Doi, M., Kimura, M.,
Komiyama, Y., Miyazaki, S., Nakata, F. et al.\ 2000b, PASJ, 52, 577 (Paper II)
\re
Akiyama, M., Ohta, K., Yamada, T., Kashikawa, N., Yagi, M.,
Kawasaki, W., Sakano, M., Tsuru, T. et al.\ 2000a, ApJ, 532, 700 (Paper I)
\re
Alighieri, S.S., Cimatti, A., \& Fosbury, R.A.E.\ 1994, ApJ, 431, 123
\re
Almaini, O., Boyle, B. J., Griffiths, R. E., Shanks, T.,
Stewart, G. C., \& Georgantopoulos, I.\ 1995, MNRAS, 277, 31
\re
Elvis, M., Wilkes, B.J., McDowell, J.C., Green, R.F., Bechtold, J.,
Willner, S.P., Oey, M.S., Polomski, E., \& Cutri, R.\ 1994, ApJS, 95, 1
\re
Fiore, F., La Franca, F., Giommi, P., Elvis, M., Matt, G.,
Comastri, A., Molendi, S., \& Gioia, I.\ 1999, MNRAS, 306, 55
\re
Francis, P.J.\ 1996, Proc. Astron. Soc. Australia, 13, 212
\re
Francis, P.J., Whiting, M.T., \& Webster, R.L.\ 2000,
Proc. Astron. Soc. Australia, 17, 56
\re
Georgantopoulos, I., Almaini, O., Shanks, T., Stewart, G.C.,
Griffiths, R.E., Boyle, B.J., \& Gunn, K.F.\ 1999, MNRAS, 305, 125
\re
Kim, D.-W., \& Elvis, M.\ 1999, ApJ, 516, 9
\re
Kobayashi, Y., Sato, S., Yamashita, T., Shiba, H., \& Takami, H.\ 1993, ApJ,
404, 94
\re
Kodama, T., \& Arimoto, N.\ 1997, A\&A, 320, 41
\re
Kodama, T., Arimoto, N., Barger, A.J., \& Arag\'on-Salamanca, A.\
1998, A\&A, 334, 99
\re
Mushotzky, R.F., Cowie, L.L., Barger, A.J., \& Arnaud, K.A.\ 2000, Nature, 404, 459
\re
Nakanishi, K., Akiyama, M., Ohta, K., \& Yamada, T.\ 2000, ApJ, 534, 587
\re
Ohta, K., Yamada, T., Nakanishi, K., Ogasaka, Y., 
Kii, T., \& Hayashida, K.\ 1996, ApJ, 458, L57
\re
Pier, E.A., \& Krolik, J.H.\ 1993, ApJ, 418, 673
\re
Serjeant, S., \& Rawlings, S.\ 1995, Nature, 379, 304
\re
Srianand, R., \& Kembhavi, A.\ 1997, ApJ, 478, 70
\re
Stocke, J.T., Morris, S.L., Gioia, I.M., Maccacaro, T.,
Schild, R., Wolter, A., Fleming, T.A., \& Henry, J.P.\
1991, ApJS, 76, 813

\begin{table*}[t]
\begin{center}
Table~1.\hspace{4pt}Summary of the $K$-band Photometry of AX~J131831+3341.\\
\end{center}
\vspace{6pt}
\begin{tabular*}{\textwidth}{@{\hspace{\tabcolsep}
\extracolsep{\fill}}lccc}
\hline\hline\\[-6pt]
                      &  $K$   & $R-K$  & $I-K$  \\
                      &  (mag) & (mag)  & (mag)    \\[4pt]\hline\\[-6pt]
 Total                & 16.38  & 3.66   & 2.58  \\
 Nucleus              & 17.48  & 4.78   & 4.29  \\
 Total $-$Nucleus     & 16.87  & 3.32   & 2.17  \\
 Region A             & 18.56  & 3.34   & 2.54  \\
 Region B             & 18.43  & 3.71   & 2.49  \\
\hline
\end{tabular*}
\end{table*}

\clearpage
\centerline{Figure Captions}
\bigskip
\begin{fv}{1}
{7cm}
{
$K$-band image of AX~J131831+3341 with an effective exposure time
of 1800 s (a). A model image (b) made by ellipse fitting and
the residual image (c) are also shown.
The field of view of the panels is $24^{\prime\prime} \times
24^{\prime\prime}$. North is up and east is to the left.
We show the surface brightness range
from $-3$ times the standard deviation to $+9$ times
the standard deviation as a gray scale with a linear scale.
Regions A and B, where the magnitudes of the extended component
were measured, are indicated with rectangles in panel a). The regions are
the same as in Paper II.
}
\end{fv}

\begin{fv}{2}
{7cm}
{
$K$-band profiles of AX~J131831+3341 (thick solid line) in
the southeastern (a) and northwestern (b) directions.
The best-fit profiles are shown by the thick dashed lines.
We plot 4 models in which the normalizations of the
nuclear component and the extended component are changed 
independently with $\pm0.2$ mag with thin solid lines.
}
\end{fv}

\begin{fv}{3}
{7cm}
{$R-I$ and $I-K$ color-color diagram of the nuclear (filled square)
and the extended (region A: filled triangle, region B: filled pentagon,
total$-$nucleus: filled circle)
components of
AX~J131831+3341.
The open squares represents the colors of optically-selected
QSOs at redshifts between 0.2 and 1.5 from Elvis et al.\ (1994).
The pentagons show the colors of the power-law model with
indices ($f_{\nu}=\nu^{\alpha}$)
of $-1.0$ (top) and $0.0$ (bottom).
The dashed line indicates the expected color range for
the $\alpha=-1$ power-law model with the
absorption derived from X-ray spectrum ($A_V = 1-6$ mag).
The absorption ($A_V > 3$ mag)  estimated
from the optical to X-ray flux ratio of the nuclear
component only permits the thick region of the line.
The tracks of the elliptical and disk models
with ages of from 0.01 Gyr to 12 Gyr are indicated by the
thick solid and dotted lines, respectively. The positions with the
models at an age of 6 Gyr are marked with tick marks on each track.
The arrow represents the effect of reddening with an $A_V$ of 1 mag.}
\end{fv}

\begin{figure}
\hspace{2cm}
\epsfbox{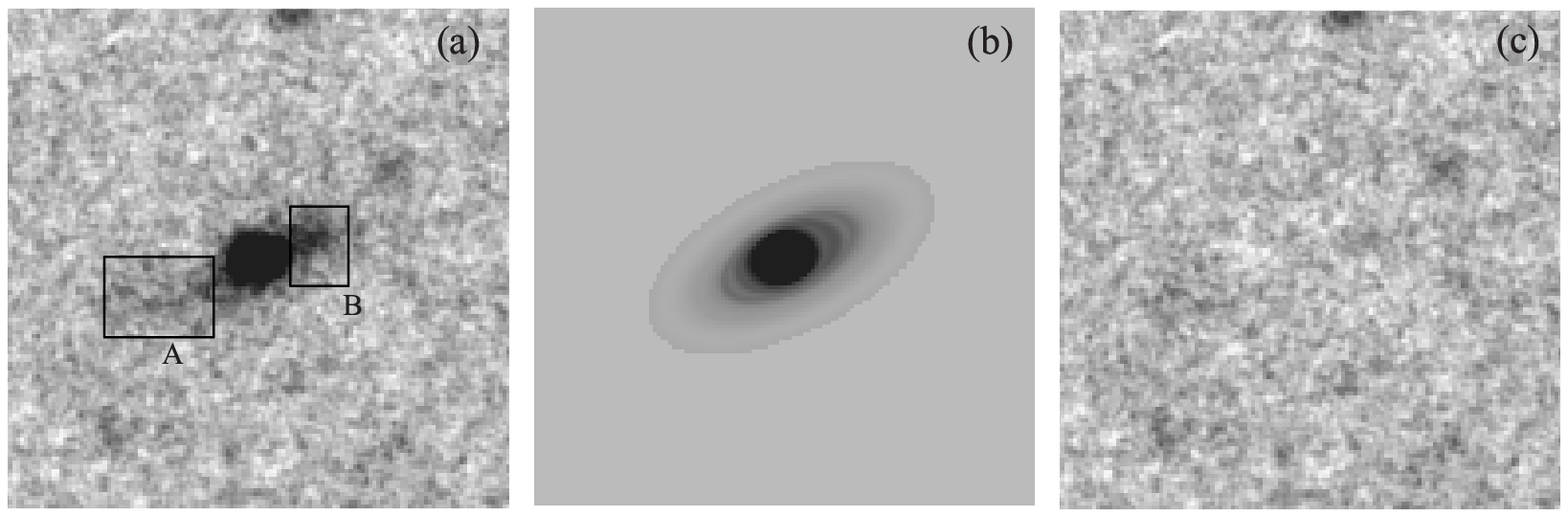} \\
\caption{}
\end{figure}

\begin{figure}
\hspace{2cm}
\epsfbox{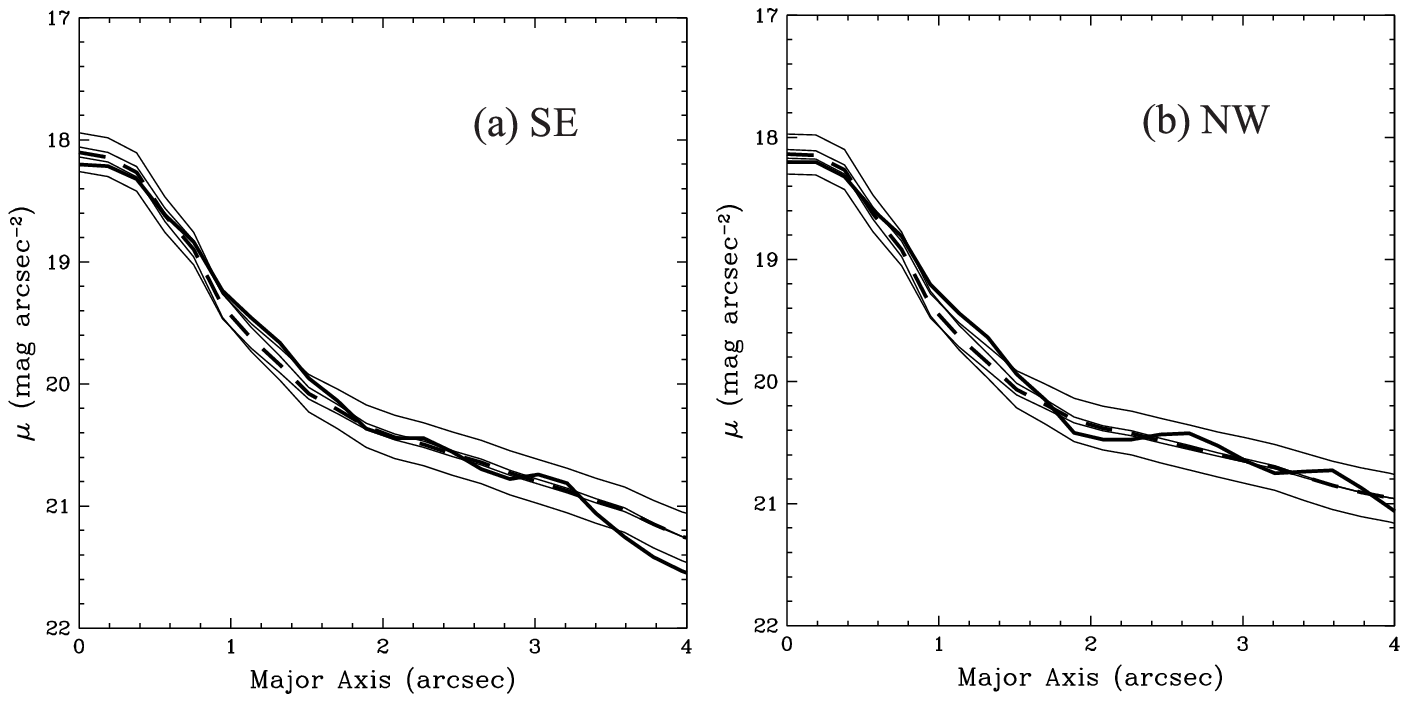} \\
\caption{
}
\end{figure}

\begin{figure}
\hspace{2cm}
\epsfbox{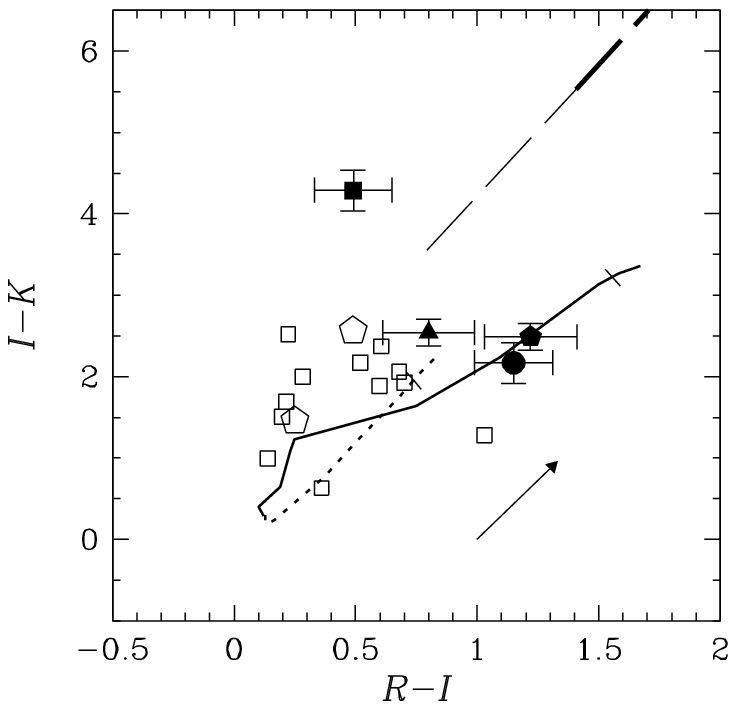} \\
\caption{}
\end{figure}

\clearpage

\end{document}